\documentclass[preprint,aps,floatfix,showpacs]{revtex4}       
\usepackage{graphicx}
\begin{document}

\title{Effects of mutual excitations in the fusion of carbon isotopes} 
\author{H. Esbensen$^1$, X. Tang$^2$, and C. L. Jiang$^1$}
\affiliation{$^1$Physics Division, Argonne National Laboratory, Argonne, 
Illinois 60439, USA,}
\affiliation{$^2$Department of Physics and JINA,
University of Notre Dame, Notre Dame, Indiana 46556, USA.}
\date{\today}

\begin{abstract}
Fusion data for 
$^{13}$C+$^{13}$C, $^{12}$C+$^{13}$C and $^{12}$C+$^{12}$C 
are analyzed by coupled-channels calculations that are based on the 
M3Y+repulsion, double-folding potential. The fusion is determined by 
ingoing-wave-boundary conditions (IWBC) that are imposed at the 
minimum of the pocket in the entrance channel potential.
Quadrupole and octupole transitions to low-lying states in projectile 
and target are included in the calculations, as well as mutual 
excitations of these states.  The effect of one-neutron transfer is 
also considered but the effect is small in the measured energy regime. 
It is shown that mutual excitations to high-lying states play a 
very important role in developing a comprehensive and consistent 
description of the measurements. 
Thus the shapes of the calculated cross sections for $^{12}$C+$^{13}$C 
and $^{13}$C+$^{13}$C are in good agreement with the data.
The fusion cross sections for $^{12}$C+$^{12}$C determined by the 
IWBC are generally larger than the measured cross sections but they 
are consistent with the maxima of some of the observed peak cross 
sections. They are therefore expected to provide an upper limit for the 
extrapolation into the low-energy regime of interest to astrophysics.
\end{abstract} 
\maketitle

\section{Introduction}

The fusion of carbon nuclei is an important reaction in the description of 
type Ia supernovae and other astronomical events in the cosmos like the 
superburst of an accreting neutron star.
There is, unfortunately, a very large uncertainty in the predicted fusion 
cross sections for $^{12}$C+$^{12}$C that are needed in a stellar 
environment \cite{gasq}, partly 
because the cross sections are difficult to measure down to the low energies 
that are of interest, and partly because the data contain strong resonance 
structures that make it difficult to extrapolate the measured cross sections 
to low energy. In order to get some constraints on the extrapolation it is  
instructive to analyze the existing fusion data for $^{13}$C+$^{13}$C by
Trentalange et al. \cite{trent} and Charvet et al. \cite{char},  
and also of the $^{12}$C+$^{13}$C data by Dayras et al. \cite{dayras} 
and Notani et al. \cite{tang1}, because these data do not exhibit strong 
resonance features. 

The $^{13}$C+$^{13}$C fusion data \cite{trent} are analyzed by 
coupled-channels calculations that include couplings to the one-phonon
quadrupole and octupole excitations in projectile and target, as well 
as mutual excitations of these states.
The effect of one-neutron transfer, which can proceed to the 
$^{12}$C+$^{14}$C mass partition with positive $Q$ value 
is also studied, and so is the effect of the neutron exchange with zero
Q value in $^{12}$C+$^{13}$C reactions. 

The fusion cross sections are determined by ingoing-wave-boundary
conditions (IWBC) that are imposed at the minimum of the pocket in the
entrance channel potential.  The calculated cross sections defined
this way are fairly smooth as functions of the center-of-mass energy
and they are well suited to analyze the fusion data for 
$^{13}$C+$^{13}$C \cite{trent,char} and $^{12}$C+$^{13}$C 
\cite{dayras,tang1}. The $^{12}$C+$^{12}$C fusion data, on the other hand,
contain a lot of structures or resonances, and it is beyond the scope
of this investigation to try to reproduce these data in detail. 
It is, however, of great interest to see how the calculated cross 
sections, obtained from the IWBC, compare to the data and how they 
possibly can be used to put constraints on the extrapolation 
to very low energies.

The coupled equations are solved using either a standard Woods-Saxon 
\cite{BW} or the M3Y+repulsion, double-folding potentials \cite{misi}. 
The coupled-channels effects on the calculated fusion cross sections 
are relatively modest compared to the large enhancement of several 
orders of magnitude that are commonly seen in calculations of heavy-ion 
fusion reactions. For the carbon systems, the enhancement of fusion
compared to the no-coupling limit is typically a factor of 2 at energies 
far below the Coulomb barrier. 
However, it is a challenge to explain the data in detail because 
the coupled-channels calculations are sensitive to channels that have
rather high excitation energies. One problem is that these channels
are closed at low beam energies but that problem can be 
solved by imposing the correct, decaying state boundary conditions 
at large distances between the reacting nuclei \cite{o16}.
Another problem is that the nuclear structure of high-lying 
states is sometimes very uncertain.

The M3Y+repulsion potential was introduced in Ref. \cite{misi}
 to explain the hindrance 
that has been observed in the fusion of many heavy-ion systems at 
very low energies \cite{sys}. The hindrance phenomenon was first observed 
as a suppression at very low energies of $^{60}$Ni+$^{89}$Y fusion data 
compared to calculations that were based on a standard Woods-Saxon 
potential \cite{niy}.
A simple explanation of the phenomenon is the existence of a shallow 
pocket in the entrance channel potential which forces the fusion 
cross section to vanish as the center-of-mass energy approaches the
minimum energy of the pocket \cite{misi}.
Important issues have been whether the fusion hindrance also occurs in 
light-ion systems, and how it will affect the extrapolation of measured 
cross section to the low energies that are of interest to astrophysics 
\cite{cljastro}.
The evidence for quasi-molecular resonances, observed in the elastic 
scattering \cite{bromleyel} and fusion reactions \cite{bromleyr,michaud} 
of $^{12}$C+$^{12}$C are usually explained as resonances 
in a shallow two-body potential between the reacting nuclei.
We will show that the analysis of the $^{13}$C+$^{13}$C and 
$^{12}$C+$^{13}$C fusion data supports the idea of a fusion hindrance
and the existence of a shallow pocket in the entrance channel potential.

The fusion cross sections reported in Refs.  
\cite{trent,char,dayras} are
based on measurements of the characteristic $\gamma$-rays emitted 
from some of the evaporation residues that are produced, mostly those 
associated with the proton, neutron, and $\alpha$ decay. 
The total fusion cross sections were obtained with the aid of 
statistical model calculations which is a major source of the 
systematic uncertainty. The systematic error 
is difficult to estimate but it can be quite large. For example, 
the systematic error of the absolute cross sections quoted in 
Ref.  \cite{trent} is $\pm$15\%. Since some of the systematic error
concerns the overall normalization of the measured cross sections,
and not so much the shape (i.~e., the energy dependence of the
cross section), it is of interest to adopt an adjustable overall
normalization when analysing the data. 

The ingredients of the coupled-channels calculations 
are presented in the next section. The analysis of the $^{13}$C+$^{13}$C
fusion data is discussed in section III, where the optimum parameters of 
the M3Y+repulsion potential, such as the radius parameter of $^{13}$C 
and the diffuseness associated with the repulsive term, are determined.
These parameters are used together with
a parameterization of the experimental density of $^{12}$C
as input to the coupled-channels calculations of the fusion cross 
sections for $^{12}$C+$^{13}$C and $^{12}$C+$^{12}$C which are 
presented in sections IV and V, respectively. 
The systematics of the low-energy fusion cross sections 
for the 3 systems of carbon isotopes is discussed in section VI.
Finally, the conclusions of this work are presented in section VII.

\section{Details of the calculations}

The ingredients in the coupled-channels calculations 
include the ion-ion potential, the nuclear structure input, the 
strength and Q-value of the transfer reactions, and the definition of
the fusion cross section. 
The coupled-channels technique that is used has been applied previously
and is described, for example, in Refs. \cite{alge,misi,opb,o16}.
All of the  details will therefore not be repeated here.
It is emphasized that the calculations are performed in the rotating frame 
approximation \cite{alge} which makes it feasible to include the most 
important reaction channels in the calculations. In this approximation, 
the magnetic quantum number $M$ of the entrance channel is conserved. 
In reactions between nuclei with $0^+$ ground states, this implies that 
the $M$ quantum number remains zero. In reactions with odd nuclei one
would have to repeat the calculations for each initial value of $M$ and 
the average fusion cross section should be compared to the measurements.
In the fusion of $^{12}$C+$^{13}$C, for example, the ground state of 
$^{12}$C is a $0^+$ state and $^{13}$C has a $1/2^-$ ground state.
However, the cross sections for $M$ = -1/2 and +1/2  are identical
so it is sufficient to calculate the cross section once.


\subsection{Densities and ion-ion potentials}

Besides the commonly used Woods-Saxon potential,
the ion-ion potential that will be used in this paper is the so-called 
M3Y+repulsion
potential \cite{misi}. It consists of the conventional M3Y potential 
and a repulsive term which is discussed below. The M3Y double-folding 
potential is defined as
\begin{equation}
U_n({\bf r}) = \int d{\bf r}_1 \ d{\bf r}_2 \
\rho_1({\bf r}_1) \ \rho_2({\bf r}_2) 
v_{M3Y}({\bf r+r_2-r_1}), 
\end{equation} 
where $v_{M3Y}(r)$ is the M3Y effective nucleon-nucleon interaction
derived from the Reid potential \cite{m3y}.
The densities $\rho_i({\bf r})$ of the $^{12}$C and $^{13}$C carbon 
isotopes 
are parametrized in terms of the symmetrized 
fermi functions defined in the appendix of Ref. \cite{opb}.
One advantage of this parametrization is that the mean square radius 
is given by the simple expression,
\begin{equation}
\langle r^2 \rangle = \frac{3}{5} \bigl(R^2 +
\frac{7}{3} \ (\pi a)^2\bigr),
\end{equation}
in terms of the radius $R$ and the diffuseness $a$.

\begin{figure}
\includegraphics[width = 8cm]{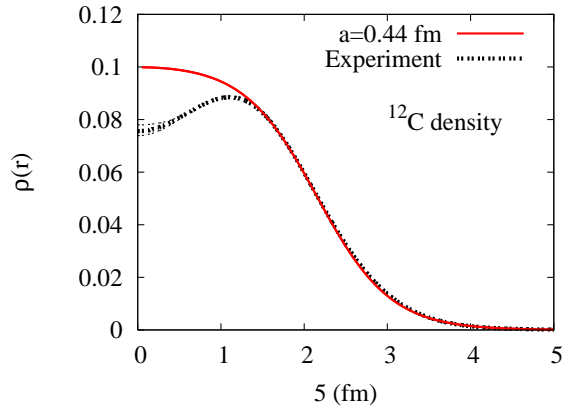}
\caption{\label{dens} 
(Color online) 
Experimental point-proton density of $^{12}$C \cite{vries} 
is compared to the symmetrized fermi function distribution with diffuseness 
$a$ = 0.44 fm and radius $R$ = 2.155 fm.}
\end{figure}

The diffuseness of the densities for both nuclei is set equal to 0.44 fm 
so the only parameters that need to be specified are the matter radii 
of the reacting nuclei. 
This choice of the diffuseness is in good agreement with the tail of
the measured point-proton density of $^{12}$C \cite{vries} which is
illustrated in Fig. \ref{dens}. The calculated point-proton distribution 
has the radius $R$ = 2.155 fm consistent with the measured rms charge radius.
The radii that reproduce the point-proton (pp) rms radii of $^{12}$C 
and $^{13}$C, extracted from the measured root-mean-square (rms) 
charge radii \cite{angeli}, are shown in the last few lines of Table I.
The quoted radius for $^{12}$C is expected to be a realistic estimate
of the matter radius of $^{12}$C and it will therefore be used in 
calculations of the M3Y double-folding potentials. 

The matter radius of $^{13}$C is possibly larger than the 
radius of $^{12}$C because of the valence neutron. That is what one
would expect by adding a neutron to an inert $^{12}$C core nucleus.
A simple estimate based on a single-particle, Woods-Saxon potential  
gives a mean square radius of $\langle r^2_n\rangle$ = 8.655 fm$^2$ 
for the $p_{1/2}$ valence neutron. Combined with the point-proton 
rms radius of $^{12}$C, 
$\langle r^2\rangle_{12}$ = 5.462(9) fm$^2$, one can now estimate 
the mean square matter radius $\langle r^2\rangle_{13}$ of $^{13}$C by
\begin{equation}
13 \langle r^2\rangle_{13} =
12 \Bigl(\langle r^2\rangle_{12} + 
\frac{\langle r^2_n\rangle}{13}\Bigr).
\label{rms}
\end{equation}
The result is a rms matter radius of 2.378 fm. The associated 
radius of a fermi-function distribution with diffuseness $a$ = 0.44 fm 
is 2.228 fm according to the last line of Table I. 
This is a useful reference value for the discussion below. 

A critical part of the M3Y+repulsion, double-folding potential is the 
repulsive part \cite{misi}, which is generated by a contact interaction,
$v_r\delta({\bf r})$. The densities that are used in calculating the
repulsive, double-folding potential have the same radius as those 
that are used in the calculation of the M3Y double-folding potential
but the diffuseness $a_r$ is chosen differently (usually smaller.)
The strength $v_r$ of the repulsion, on the other hand, is adjusted
for a given value of $a_r$ so that the nuclear incompressibility 
$K$ = 234 MeV is produced (see Ref. \cite{misi} for details.) 
With this constraint there are three parameters that must be 
specified before the M3Y+repulsion potential can be calculated, 
namely, the matter radii of the two reacting nuclei, and the 
diffuseness parameter $a_r$ associated with the repulsion. 

The parameters that give the best fit to Trentalange's $^{13}$C+$^{13}$C 
fusion data are the $^{13}$C radius 
$R$ = 2.28 fm and the diffuseness $a_r$ = 0.31 fm.  
How these parameters were determined is described
in detail in section III.B. The radius $R$ = 2.28 fm is slightly 
larger than the 2.228 fm that was estimated from Eq. (\ref{rms}). 
That may not be unreasonable as discussed in Ref. \cite{4848} 
because using a larger radius in a coupled-channels calculation is a 
way to compensate for the dynamic polarization of states that are not 
included explicitly in the calculation. 
The discrepancy between the estimated rms radius, Eq. (\ref{rms}), and
the value extracted from the fit to the fusion data is only 1.2 \%.

\begin{figure}
\includegraphics[width = 17cm]{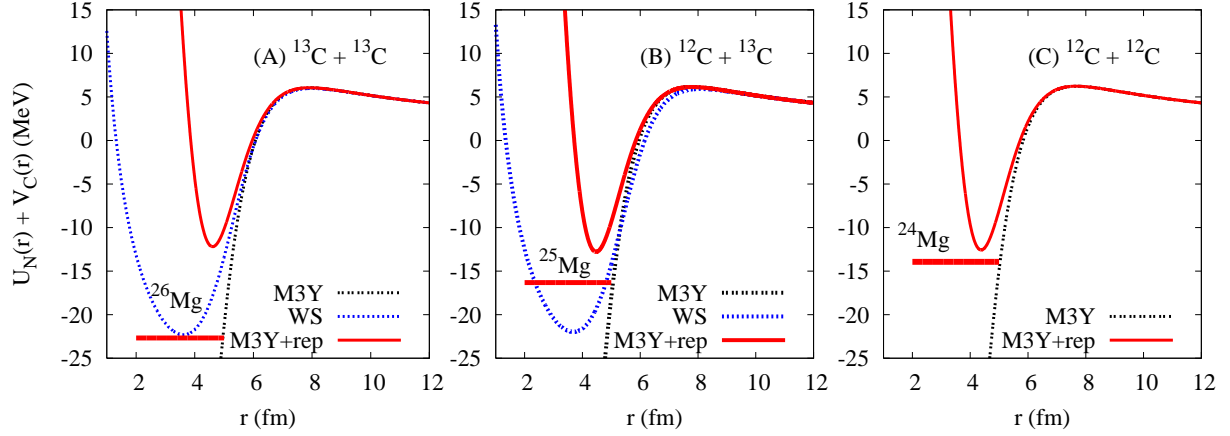}
\caption{\label{pot1313} 
(Color online) 
The M3Y+repulsion double-folding potentials
for different combinations of carbon isotopes.
They are based on the parameter $a_r$ = 0.31 fm and the radius 
$R$ = 2.28 fm for $^{13}$C, and $R$=2.155 fm for $^{12}$C.
The black dashed cures are the pure M3Y potentials.
Also shown are the WS potentials discussed in the text and the 
ground state energies of the compound nuclei.}
\end{figure}

The entrance channel potential determined by the best fit parameters
is shown by the solid curve in Fig. \ref{pot1313}A. Also shown in 
Fig. \ref{pot1313} are the potentials that will be applied in
calculating the fusion of $^{12}$C+$^{13}$C and $^{12}$C+$^{12}$C. 
They were obtained by setting the radius of $^{12}$C equal to the 
radius $R$ = 2.155 fm of the point-proton distribution of $^{12}$C 
(see Table I), and the diffuseness associated 
with the repulsion was set to $a_r$ = 0.31 fm.

The pure M3Y double-folding potentials shown in Fig. \ref{pot1313} are
unphysical because they are much deeper for overlapping nuclei than the 
ground state energy of the compound nucleus. 
The (blue) dashed curve in Fig. \ref{pot1313}A is the entrance channel 
potential for a standard Woods-Saxon (WS) potential with the depth $V_0$ = 
-39.31 MeV, radius $R_{ws}$ = 5.268 fm, and diffuseness $a$ = 0.63 fm. 
Here the radius was adjusted to optimize the fit to Trentalange's 
fusion data as discussed in section III.A. This potential is deeper 
than the M3Y+repulsion potential and has a minimum that is close to 
the energy of the compound nucleus $^{26}$Mg. 

The three entrance channel potentials shown in Fig. \ref{pot1313}A,  
which are based on the M3Y, the M3Y+repulsion and the Woods-Saxon
potential, produce essentially the same Coulomb barrier height.
$V_{CB}\approx$ 6 MeV. 
The most important features of the M3Y+repulsion potential are that 
it produces a shallow pocket and a relatively thick barrier. 
These features were utilized in Ref. \cite{misi} to explain the 
hindrance of fusion that has been observed in the fusion of many 
medium mass systems at extreme subbarrier energies.
The analysis of the $^{13}$C+$^{13}$C fusion data discussed in the
next section confirms that a shallow pocket in the entrance channel 
is a general feature that helps explain the energy dependence of 
fusion data.

\subsection{Nuclear couplings}

The nuclear couplings that excite the low-lying states of 
projectile and target are derived from the macroscopic description 
of surface excitations \cite{BW}. In this description, the 
surface of nucleus $i$ is parametrized as
\begin{equation}
R_i({\hat r}) = R_i^{(0)} \Bigl(1 + \sum \alpha_{\lambda\mu}^{(i)} 
Y_{\lambda\mu}^*({\hat r})\Bigr),
\end{equation}
where $\alpha_{\lambda\mu}^{(i)}$ are the (static or dynamic)
deformation amplitudes of the nucleus, and ${\hat r}$ specifies 
a spatial direction.  The distortion of the nuclear surface,
\begin{equation}
\delta R_i = R_i^{(0)} \sum \alpha_{\lambda\mu}^{(i)} 
Y_{\lambda\mu}^*({\hat r}),
\end{equation}
is an operator that can excite and de-excite the nucleus
(see chapter II.4 of Ref. \cite{BW}.)
The excitation can be vibrational, in which case the intrinsic
hamiltonian is a harmonic oscillator, or it can be rotational,
in which case $\alpha_{\lambda\mu}$ = $\beta_\lambda$
$D^\lambda_{\mu 0}({\hat e})$, where $\beta_\lambda$ is the
static deformation parameter and ${\hat e}$ is the direction 
of the symmetry axis of the deformed nucleus.

In the macroscopic description of heavy-ion reactions \cite{BW}, 
the nuclear potential between two colliding nuclei is parametrized 
in terms of the surface distortions $\delta R_i$ as follows
\begin{equation}
V_N(r-R_1^{(0)}-R_2^{(0)}-\delta R_1({\hat r})-\delta R_2(-{\hat r})),
\end{equation}
where ${\hat r}$ is the direction of the center of mass distance 
between projectile and target.
In the rotating frame approximation, which is a simplifications that
is commonly used and which will be used in coupled-channels calculations, 
the direction of ${\hat r}$ defines the z-axis, so that
\begin{equation}
\delta R_1({\hat r}) = R_1^{(0)} \sum_\lambda 
\alpha_{\lambda 0}^{(1)} \ \sqrt{\frac{2\lambda+1}{4\pi}}.
\end{equation} 

In the model of heavy-ion fusion reactions described 
in Ref. \cite{alge}, the nuclear potential is expanded up to
second order in the surface distortions,
$$U_N(r,\delta R_1,\delta R_2) = V_N 
- \frac{dV_N}{dr}(\delta R_1+\delta R_2)$$
\begin{equation}
+ \frac{1}{2} \ \frac{d^2V_N}{dr^2} \
\Bigl[(\delta R_1+\delta R_2)^2 -
\langle 0| (\delta R_1+\delta R_2)^2 | 0\rangle\Bigr],
\label{uexp}
\end{equation}
where the argument of $V_N$ and its derivatives is
$r-R_1^{(0)}-R_2^{(0)}$.
The second-order term in this expression has been
renormalized so that the ground state expectation value 
of that term is zero.  
The ground state expectation value of the first-order term 
(proportional to $\delta R_1$+$\delta R_2$) will also vanish 
if the ground states of the two reacting nuclei are $0^+$ states.
The ground state expectation of the entire expression, 
Eq. (\ref{uexp}), is therefore identical to the `bare' interaction, 
i.~e.,
\begin{equation}
\langle 0| U_N(r,\delta R_1,\delta R_2) | 0\rangle =
V_N(r-R_1^{(0)}-R_2^{(0)}).
\end{equation}
This implies that 
the M3Y+repulsion potential which is calculated using the
ground state densities of the reacting nuclei  
can be used as the bare interaction, $V_N$.

The diagonal matrix element of the first-order term in Eq. 
(\ref{uexp}) can be non-zero, for example, in the $2^+$ state 
of a deformed nucleus. The calculation of diagonal and
off-diagonal matrix elements is outlined in the appendix.
Details of how to calculate the matrix elements of the 
second-order couplings in Eq. (\ref{uexp}), both for vibrational
and rotational excitations, are given in Ref. \cite{alge}.

\subsection{Nuclear structure input}

The nuclear structure input 
to the coupled-channels calculations is listed in Table II. 
The nuclei $^{12}$C and $^{13}$C are both considered deformed 
with oblate quadrupole shapes. The connection between the deformation
parameter $\beta_\lambda$  and the measured $B(E2)$ values is summarized 
in the appendix.  For $^{12}$C one can extract the quadrupole deformation 
parameter $\beta_2$ = 0.57(2) from the measured strength of the 
$0_1^+\rightarrow 2^+$ transition. 
The value of this deformation parameter produces an intrinsic quadrupole 
moment of $Q_0$ = -19.5 fm$^2$ (assuming an oblate shape). The quadrupole 
moment of the $2^+$ state is therefore $Q_2$ = -2/7$Q_0$ = 5.57 fm$^2$, 
which is consistent with the measured value of $6\pm 3$ fm$^2$ \cite{vermeer}.
Also shown in the Table is the structure information about the 
$2^+\rightarrow0_2^+$ transition which is included in the calculations
as part of the two-phonon quadrupole excitation of $^{12}$C.

Values of quadrupole deformation parameter $\beta_2$ of $^{13}$C, 
extracted from Eq. (\ref{belwu}) of the appendix and the known 
strengths of the quadrupole transition from the 1/2$^-$ ground to 
the 3/2$^-$ and the 5/2$^-$ excited states, are shown in the last 
column of Table II. They are seen to be almost identical, which 
indicates that $^{13}$C is a fairly good rotor. Assuming the 
deformation parameter $\beta_2$ = 0.495 and an oblate shape, 
one obtains the intrinsic quadrupole moment $Q_0$ = -17.9 fm$^2$.

The strength of the octupole transition in $^{13}$C quoted in Table II
appears to be very large and almost as large as the octupole strength 
in $^{12}$C. This is misleading because what matters is the off-diagonal 
matrix element of the octupole amplitude, which is given by Eq. 
(\ref{matel}) of the appendix.  In $^{12}$C the matrix element is
\begin{equation}
\langle 3 0 0 |\alpha_{30}| 0 0 0 \rangle = \frac{\beta_3}{\sqrt{7}},
\label{octc12}
\end{equation} 
whereas in $^{13}$C it is (with $M$ = $K$= 1/2)
\begin{equation}
\langle{\frac{5}{2} \ \frac{1}{2} \ \frac{1}{2}} |\alpha_{30}|
\frac{1}{2} \ \frac{1}{2} \ \frac{1}{2}\rangle =
\beta_3 \sqrt{\frac{2}{6}} \ 
\langle \frac{1}{2} \frac{1}{2} \ 3 0 | \frac{5}{2} \frac{1}{2}\rangle^2 
= \sqrt{\frac{3}{7}} \ \frac{\beta_3}{\sqrt{7}}.
\label{octupole}
\end{equation}
The matrix element of the octupole amplitude is therefore reduced  
in $^{13}$C by the factor $\sqrt{3/7}\approx$ 0.655 relative to the 
expression Eq. (\ref{octc12}) for $^{12}$C. 

The coupled-channels calculations that are performed are similar to 
those presented in Ref. \cite{alge} for the fusion of $^{27}$Al with 
different germanium isotopes. In addition to the nuclear couplings 
derived from the nuclear interaction, Eq. (\ref{uexp}), the Coulomb 
interaction (Eq. (2) of Ref. \cite{alge}) is included in the 
calculations to first order in the 
deformation amplitudes.

\subsection{Transfer reactions}

The one-neutron transfer reactions that will be considered all 
involve the $p_{1/2}$ orbit, both in the initial and final states. 
One example is the ground state to ground state transfer reaction, 
$^{12}$C($^{13}$C,$^{12}$C)$^{13}$C, which has zero Q value.
The spectroscopic factors for the initial and final state $1/2^-$ 
state in $^{13}$C were set equal to 0.75(10), which is the 
value recommended in Table III of Ref. \cite{jenny}.

The other example is the one-neutron transfer from $^{13}$C+$^{13}$C to 
the ground states of $^{12}$C+$^{14}$C which has a Q-value of +3.2 MeV.
The spectroscopic factor for the initial state is the same as above.
It is set equal to 2*0.75 in the calculations because the 
transfer can take place from either of the two $^{13}$C nuclei in the 
entrance channel. 
The spectroscopic factor for the $p_{1/2}$ orbit in the $0^+$ ground 
state of $^{14}$C is 1.63(33) according to table III of Ref. \cite{jenny}.

The transfer form factors that are used are the so-called Quesada 
single-particle form factors \cite{quesada} which in the past  
turned out to be fairly realistic \cite{opb,NPA492}. 
They have here been calibrated against calculations performed with the 
computer code Ptolemy \cite{ptol} of the one-neutron transfer cross in 
$^{12}$C+$^{13}$C collisions at a 5.2 MeV center-of-mass energy.


\subsection{Calculation of fusion cross sections}

The fusion cross section is calculated from the ingoing flux which
is determined by the ingoing-wave-boundary conditions (IWBC) that
are imposed at the minimum of the pocket in the entrance channel
potential. This approximation ignores the internal structure of
the combined di-nuclear or compound system and the calculated 
cross sections are usually fairly smooth functions of energies,
at least at energies near and below the Coulomb barrier.

At energies far above the Coulomb barrier, there are sometimes 
numerical problems in the coupled-channels calculations which can 
cause an erratic behavior of the calculated cross section as a
function of energy. One can overcome these problems by applying 
a weak imaginary potential \cite{misi}
\begin{equation}
W(r) = \frac{W_0}{1 + \exp((r-R_{\rm w})/a_{\rm w})},
\end{equation}
which acts near the position $R_{\rm w}$ of the pocket in the entrance 
channel potential. The fusion cross section is then defined as the
sum of the absorption cross section and the cross section for the
ingoing flux.
A weak imaginary potential will be applied in the following when 
the calculated fusion cross section is shown in a linear plot at 
energies far above the Coulomb barrier. The parameters that are used 
are $W_0$ = -5 MeV and $a_{\rm w}$ = 0.5 fm.  No imaginary potential 
will be used in the calculations at energies below the Coulomb barrier. 

The 1/2$^-$ ground state spin of $^{13}$C causes some concern when 
calculating the scattering and fusion of $^{13}$C+$^{13}$C. 
The fermionic nature of $^{13}$C requires that the total wave 
function for $^{13}$C+$^{13}$C be anti-symmetric. By coupling the 
intrinsic $1/2^-$ ground state spins of the reacting nuclei one 
obtains a total spin of $S$=0 (anti-symmetric) and $S$=1 (symmetric). 
The wave function for the relative motion of the two nuclei must 
therefore be symmetric for $S$=0, i.~e., it consists of even partial 
waves, and it must be antisymmetric for $S$=1, consisting of odd 
partial waves. The fusion cross section is therefore calculated 
as the weighted sum
\begin{equation}
\sigma_f =
\frac{1}{2}\ \sigma({\rm even} \ L) +   
\frac{3}{2}\ \sigma({\rm odd} \ L).
\label{fusa}
\end{equation}

It turns out that the contributions to the fusion cross section
from odd and even $L$-values are almost identical.  The biased 
weighting of the two contributions in Eq. (\ref{fusa}) does 
therefore not have much effect.
In calculations of the fusion and elastic scattering of two 
identical spin zero $^{12}$C nuclei, one should only consider 
the contribution from even partial waves which is doubled 
because of the symmetry of two identical particles.
In reactions of $^{12}$C+$^{13}$C, on the other hand, the 
contribution from even and odd partial waves have the same weight.

\section{Analysis of the $^{13}$C+$^{13}$C fusion data} 

\begin{figure}
\includegraphics[width = 10cm]{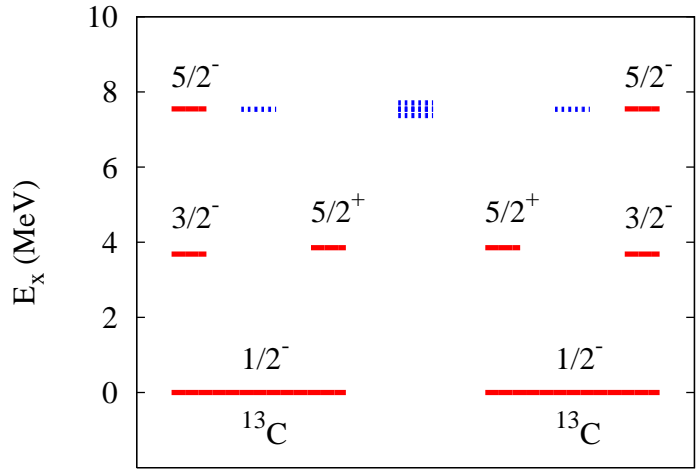}
\caption{\label{spc1313} 
(Color online) 
Excitation energies of the channels considered in the calculations of 
the fusion of $^{13}$C+$^{13}$C.
The (red) solid lines are the excitations in projectile
and target of the $^{13}$C states shown in Table I.  
The (blue) dashed lines are the 6 mutual excitations 
that involve the 3/2$^-$ and 5/2$^+$ states in 
projectile and target.}
\end{figure}

In this section the $^{13}$C+$^{13}$C fusion data by Trentalange 
et al. \cite{trent} and Charvet et al. \cite{char} are analyzed 
in terms of coupled-channels calculations. The channels that are 
included are enumerated in Fig. \ref{spc1313}. The (red) solid 
lines represent the elastic channel and the channels associated 
with the excitation of the states shown in Table II, in both 
projectile or target. The calculation based on these channels 
is referred to as the one-phonon (ph1) calculation and has a 
total of 7 channels.

The (blue) dashed lines in Fig. \ref{spc1313} are the 6 mutual 
excitations of the lowest quadrupole and octupole excitations 
in projectile and target. These include the four mutual  states 
$(I_1,I_2)$, where $I_i$ = $3/2^-$ or $5/2^+$, $i$=1,2, and $I_1$ 
and $I_2$ belong to different nuclei, and the two $(3/2^-,5/2^+)$ 
mutual excitations  that belong to the same nucleus.
Together with the seven channels of the one-phonon calculation
that gives a total of 13 channels. This calculation is referred 
to as the mutual excitation calculation or Ch13. 

The calculation Ch13 will be compared to the one-phonon calculation 
(ph1) described above and the no-coupling limit which has 
only one channel. There are other excitations, for example, the 
two-phonon excitations which are poorly known and mutual 
excitations that involve the $5/2^-$ state but they are all 
ignored in the following.  

It can be very difficult to see small discrepancies between measured 
and calculated cross sections, in particular, when they are plotted 
on a conventional logarithmic scale. It is therefore useful 
to use other representations that amplify certain features of 
the comparison, such as the $S$ factor for fusion, which emphasizes
the behavior at low energies. Another representation, which is
very useful when coupled-channels effects are modest, is the 
so-called enhancement factor, which was used in the early 
days when the enhancement of subbarrier fusion was first discovered. 
It is defined as the ratio of a cross section relative to the 
cross section calculated in the no-coupling limit, and it
will be used both for measurements as well as coupled-channels 
calculations, as long as it is indicated which no-coupling 
limit is used as a reference.
Finally, the comparison of data and coupled-channels calculations
will also be made in terms of the ratio of the measured and calculated 
cross sections. If the measured and calculated cross sections agree, 
the ratio should be close to one. 

The one-neutron transfer discussed earlier is included as 
an independent degree of freedom. That implies that the transfer 
can take place from any excited state and it is calculated with
the same form factor which describes the ground state to ground 
state transition.
The full calculation (with 13 excitation channels mentioned 
above and the one-neutron transfer) will therefore contain 
26 channels and is referred to as the Ch26 calculation.

The analysis of the data is performed in terms of the $\chi^2$ 
per data point, $\chi^2/N$, which is minimized by adjusting certain 
parameters of the ion-ion potential. The overall normalization of 
the calculated cross sections is also adjusted by a 
multiplicative factor $S_c$, in order to improve the fit to the data.
The motivation for this adjustment is the large uncertainty in the
absolute normalization of the measured cross sections that was 
mentioned in the introduction.  The shape (i.e., the energy-dependence) 
of the measured cross section is assumed to be more accurately
determined. The uncertainty in the data analysis will therefore 
include the statistical uncertainty and an adopted systematic error 
of only 3\%, which is much smaller than the 15\% 
experimental error \cite{trent}. 

\subsection{Application of Woods-Saxon potential}

\begin{figure}
\includegraphics[width = 8cm]{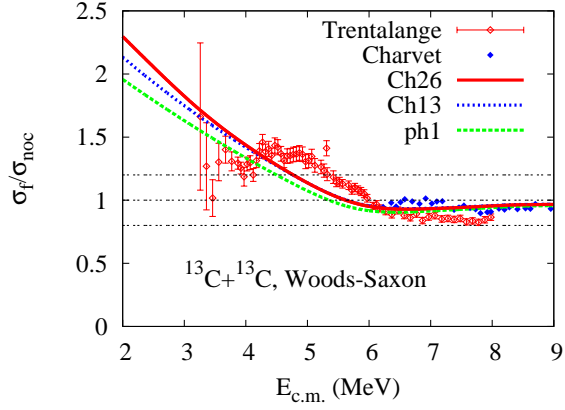}
\caption{\label{frat1313ws} 
(Color online) 
Enhancement of the fusion of $^{13}$C+$^{13}$C 
relative to the cross section of the no-coupling calculation.
Shown are the enhancement of the one-phonon calculation (ph1), 
the mutual excitation (Ch13) calculation, and the full calculation
(Ch26) which also includes one-neutron transfer. 
The calculations are based on a standard Woods-Saxon potential
with radius $R_{ws}$ = 5.268 fm. Also shown are the ratios of the measured 
cross sections \cite{trent,char} relative to the no-coupling limit.}
\end{figure}

The first set of calculations are based on a standard Woods-Saxon 
potential with typical parameters from Ref. \cite{BW}. 
The diffuseness of the potential is set to $a$ = 0.63 fm and the 
depth is -39.31 MeV. The radius of the potential, $R_{ws}$ = 5.268 fm,
was adjusted to optimize the fit to the data of Ref. \cite{trent}.
A comparison to the data is made in Fig. \ref{frat1313ws} in 
terms of the enhancement of the cross sections with respect to 
the no-coupling limit. The enhancement of the measured cross sections
is seen to deviate from the enhancement of the coupled-channels 
calculation (Ch26) by up to 20\%. Moreover, the effects of mutual 
excitations and one-neutron transfer are seen to be modest by
comparing to the no-transfer calculations (Ch13) and the 
one-phonon calculations (ph1).
It is very unlikely that the agreement with the data can 
be improved much by adjusting the strengths of the couplings to the 
excitation or transfer channels. It is shown in the next subsection 
that a much better agreement with the data is achieved by applying 
the M3Y+repulsion instead of the Woods-Saxon potential. 

The agreement with the data in Fig. \ref{frat1313ws} cannot be 
improved much by scaling the calculated cross sections with an
adjustable scaling factor $S_c$ in the data analysis.  
The best fit for $S_c$ = 1 has a $\chi^2/N$ = 8.7.  
It can be reduced to $\chi^2/N$ = 6.8 by adjusting both the scaling 
factor $S_c$ and the radius $R_{ws}$ of the Woods-Saxon well.
The values that give the  best fit are $S_c$ = 0.885 and 
$R_{ws}$ = 5.345 fm.

Another observation in Fig. \ref{frat1313ws} is that the data sets 
of Trentalange et al.  \cite{trent} and Charvet et al. \cite{char} 
differ by up to 10\% in the overlapping energy regime. However,
this is not a serious problem because the systematic error of both
experiments is about 15\%.



\subsection{Application of the M3Y+repulsion potential}

One advantage of the M3Y+repulsion potential is that it has an 
additional adjustable parameter besides the radius $R$ of the density, 
namely, the diffuseness parameter $a_r$ of the density that is used 
in calculating the repulsive part of the potential.
This parameter controls the depth of the pocket and the thickness 
of the barrier in the entrance channel potential.

\begin{figure}
\includegraphics[width = 10cm]{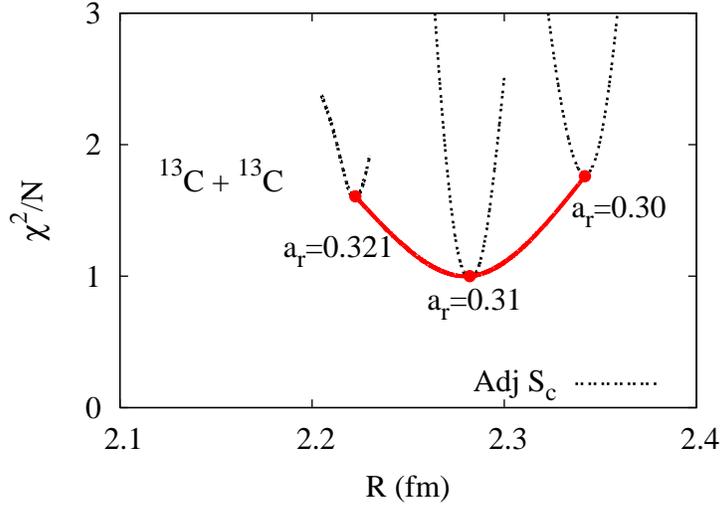}
\caption{\label{xki2} 
(Color online) 
Results of the $\chi^2$ analysis of the $^{13}$C+$^{13}$C fusion 
data \cite{trent} as function of the radius of $^{13}$C.
The dashed curves show the $\chi^2/N$ for the indicated fixed 
values of $a_r$ and minimized with respect to the scaling factor 
$S_c$. The parameters of the best solution for $a_r$ = 0.31 fm
are shown in Table 1.}
\end{figure}

The determination of the best fit parameters is illustrated 
in Fig. \ref{xki2}. For a fixed value of $a_r$, the fusion cross 
sections were calculated for different values of the radius $R$ of 
$^{13}$C. In each case the scaling factor $S_c$ was adjusted to 
give the best $\chi^2/N$. The results are shown by the dashed 
curves. The minimum $\chi^2/N$ for each dashed curve defines the
solid curve, and the minimum of this curve defines the absolute 
best fit (ABF) to the data. The parameters of the minimum are 
$a_r$ = 0.31 fm, the matter radius $R$ = 2.28 fm, and the scaling 
factor $S_c$ = 0.843. They are compared in Table 1 to other parameters
that were obtained for fixed choices of the radius of $^{13}$C.

The $\chi^2/N$ of the best fit to Trentalange's data is small
according to Table 1, $\chi^2/N$ =1.0,  although the adopted 
systematic error of the analysis was set to only 3\%. 
The scaling factor of the best fit, $S_c$ = 0.843, is not unreasonable
but it is slightly extreme because the systematic uncertainty in the
absolute cross section of the measurement was estimated to be
15\% \cite{trent}. It is therefore of interest to discuss whether 
the other parameters of the best fit are realistic.
The diffuseness $a_r$ associated with the repulsion, $a_r$ = 0.31 fm, 
falls below the range of values $a_r$ = 0.4 -- 0.43 fm that have 
been obtained in the analysis of fusion data for other symmetric 
heavy-ion systems \cite{4848}. 

The $^{13}$C radius of the best fit, $R$ = 2.28 fm, produces 
an rms matter radius of 2.407 fm according to Table 1 which 
is slightly larger than the rms matter radius of 2.378 fm 
that was estimated in Eq. (\ref{rms}).  This result is not unreasonable
as discussed in connection with an analysis of the fusion 
data for $^{48}$Ca+$^{48}$Ca \cite{4848}. 
There it was pointed out that a larger radius may simulate the 
effect of the dynamic polarization of excited states which are 
not included explicitly in the coupled-channels calculations.

If the scaling parameter $S_c$ is kept fixed at $S_c$=1 in the 
data analysis, the best fit is achieved with a smaller 
$^{13}$C radius, namely, $R$ = 2.17 fm. This value is almost
as small as the radius for the point-proton density distribution 
of $^{13}$C.  The fit is poor, with a $\chi^2/N$ = 2.75 as 
shown in the first line of Table I. 
The value of the diffuseness parameter $a_r$ which produces the 
best fit in this case is $a_r$ = 0.33.

\begin{figure}
\includegraphics[width = 10cm]{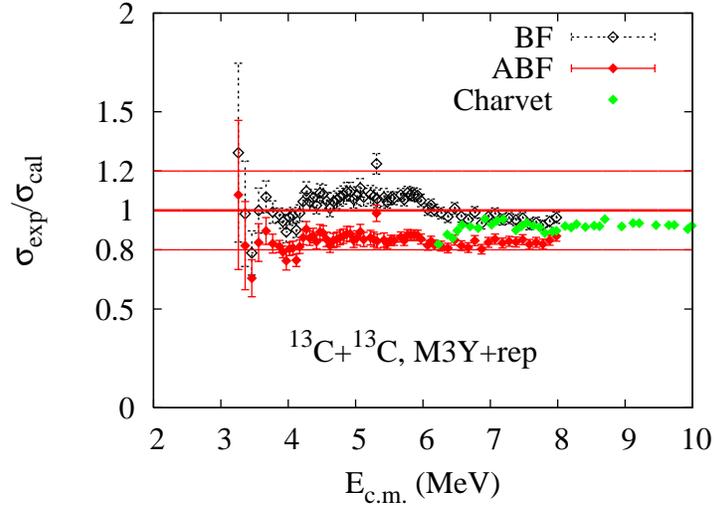}
\caption{\label{frf1313} 
(Color online) 
Ratios of the measured \cite{trent} and calculated fusion cross 
sections for $^{13}$C+$^{13}$C.
The coupled-channels calculations include 26 channels 
and are based on the M3Y+repulsion potential using the $^{13}$C 
radius $R$ = 2.17 fm (BF, $S_c$=1) and $R$=2.28 fm 
(ABF, $S_c$=0.843), respectively.
Also shown is the ratio of Charvet's data \cite{char} and the
calculation that uses the radius $R$ = 2.28 fm.}
\end{figure}

The ratios of the measured and calculated cross sections discussed 
above are illustrated in Fig. \ref{frf1313}.
The ratio of Trentalange's data \cite{trent} to the best fit (BF)
solution with radius $R$ = 2.17 and $S_c$ = 1 is shown by the open symbols. 
The ratio of the same data set to the absolute best solution (ABF) 
with radius $R$ = 2.28 fm is shown by the solid (red) points.  
Also shown at higher energies is the ratio of Charvet's data \cite{char} 
to the same solution (with radius $R$ = 2.28 fm.) 
It is seen that the ratios for the absolute best solution are essentially 
constants ($\approx$ 0.84 for Trentalange's data, and $\approx$ 0.92 
for Chatvet's data up to 10 MeV).
The two data sets differ by up to 10\% in the overlapping energy regime.
However, this is not a serious problem as mentioned earlier because the 
systematic error of both experiments is about 15\%.

The ratio for best fit (BF) solution for $S_c$=1 exhibits some energy 
dependence in Fig. \ref{frf1313}. It is of course possible that the 
data do contain some structures (similar to those observed 
in the $^{12}$C+$^{12}$C fusion data) that should not be accounted for 
by the type of coupled-channels calculations performed here. 
The absolute best fit (ABF) solution, obtained with the $^{13}$C 
radius $R$ = 2.28 fm and $S_c$ = 0.843, could therefore be misguided
or misleading.  On the other hand, the absolute best fit does have 
some attractive features as argued above: the quality of the fit is 
excellent, and the 
extracted rms matter radius is only 1.2\% larger than expected.
The absolute best fit requires a scaling of the calculation,
$S_c$ = 0.843, which is not unreasonable considering the large 
15\%  systematic error of the experiment.
The parameters of the ABF are therefore adopted in the following. 


\subsection{Results of the analysis}

\begin{figure}
\includegraphics[width = 10cm]{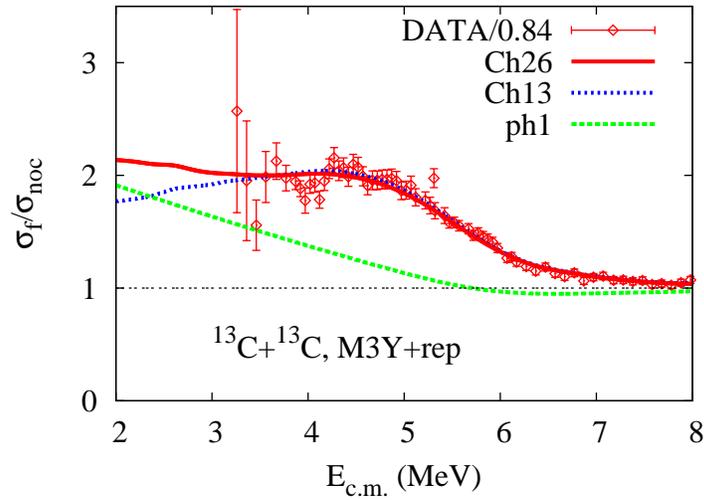}
\caption{\label{frat1313} 
(Color online) 
Enhancement of the fusion of $^{13}$C+$^{13}$C \cite{trent} 
relative to the cross sections obtained in the no-coupling limit.
The curves show the enhancement of calculations that include one-phonon 
(ph1) and mutual excitations (Ch13), and the full calculation (Ch26)
which also includes one-neutron transfer. 
The enhancement factors for Trentalange's data \cite{trent} have been 
divided by the optimum scaling factor, $S_c$ = 0.843.}
\end{figure}

In order to illustrate what makes it possible to reproduce the 
energy dependence of Trentalange's data so well when using 
the M3Y+repulsion potential, we show in Fig. \ref{frat1313} 
the enhancement of the calculations and the data relative to 
the no-coupling calculation.
In contrast to the results for the Woods-Saxon potential shown 
in Fig. \ref{frat1313ws}, there is now a strong sensitivity 
to mutual excitations. This is caused by a larger second derivative 
of the ion-ion potential (stemming from the onset of the strong
repulsion), and consequently, a stronger quadratic coupling of the 
entrance channel to mutual excitations.

An interesting observation in Fig. \ref{frat1313} is that the neutron 
transfer does not have much influence on the calculated cross section 
in the energy regime of the experiment. This is seen by comparing the
calculations labeled Ch13 and Ch26. 
The influence shows up at energies below 3 MeV, so it will have an 
impact on the extrapolation of the cross sections to lower energies. 
Another interesting feature in Fig. \ref{frat1313} is the structure 
of the data at low energies. The small peaks at 3.6 and 4.2 MeV, for 
example, could be remnants of similar structures observed in the 
fusion data for $^{12}$C+$^{12}$C.

\begin{figure}
\includegraphics[width = 10cm]{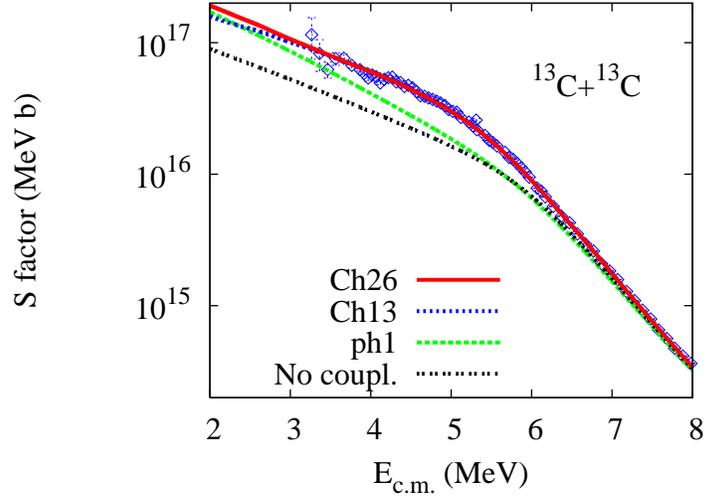}
\caption{\label{fsf1313}
(Color online) 
$S$ factors for the fusion of $^{13}$C+$^{13}$C 
calculated using the M3Y+repulsion potential.
They are compared to the data of Ref. \cite{trent}, 
which have been divided by the scaling factor $S_c$ = 0.843.}
\end{figure}

The calculated $S$ factors for the fusion of $^{13}$C+$^{13}$C are
compared in Fig. \ref{fsf1313} to the data of Ref. \cite{trent}.
The data have here been divided by the scaling factor $S_c$ = 0.843
which optimizes the agreement with the full calculation labeled Ch26.
The calculation that is based on one-phonon excitations only (ph1) 
does no reproduce the energy dependence of the data between 3 and
6 MeV. The calculation that includes mutual excitations but no 
transfer (the Ch13 calculation) is seen to reproduce the corrected 
data very well.  Finally, the additional coupling to the one-neutron 
transfer produces some enhancement in the full Ch26 calculation 
but that occurs at energies below the range of the experiment.

\begin{figure}
\includegraphics[width = 10cm]{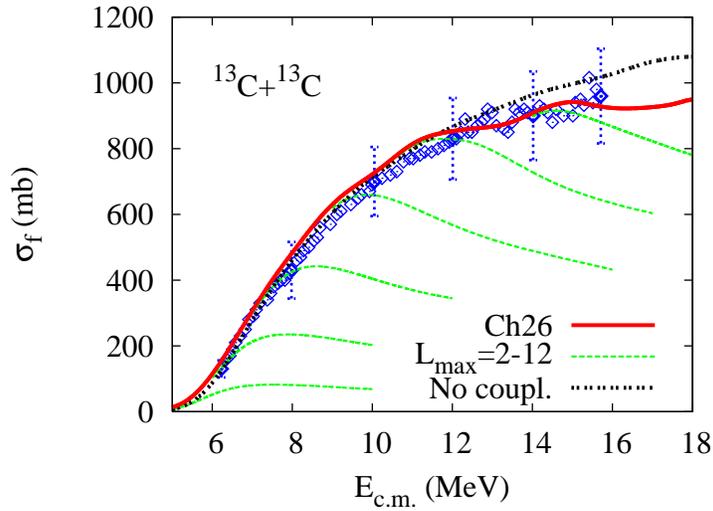}
\caption{\label{flf1313}
(Color online) 
Measured fusion cross sections for $^{13}$C+$^{13}$C \cite{char} 
are compared to the full coupled-channels calculation 
(Ch26, the solid curve).  The thin (green) dashed curves are the 
calculated cross sections for a maximum angular momentum of $L_{max}$ = 
2, 4, 6, 8, 10 and 12. The (black) dashed curve shows the 
no-coupling limit.}
\end{figure}

The calculated cross sections obtained with all 26 coupled channels 
and the parameters determined in the previous subsection are compared 
in Fig. \ref{flf1313} to the data of Ref. \cite{char}. 
%
The calculation exceeds 
the data on average by only 4\%, which is small compared to the 15\% 
systematic error of the experiment. The coupled-channels effects are 
fairly modest at high energies; this can be seen by comparing to the 
no-coupling calculation which is shown by the thick dashed curve. 
The results of a maximum angular momentum cutoff in the coupled-channels 
calculations are shown for $L_{max}$ = 2 - 12. 


\section{Predictions of the fusion of $^{12}$C+$^{13}$C}

Having determined the radius of $^{13}$C and the diffuseness $a_r$
associated with the repulsive part of the double-folding potential 
in the previous section, one can now predict the ion-ion potential
for $^{12}$C+$^{13}$C, provided the density of $^{12}$C is known.
Here the parameters of the point-proton density of $^{12}$C given 
in Table 1 are adopted. The predicted M3Y+repulsion potential is 
shown in Fig. \ref{pot1313}B. It is similar to $^{13}$C+$^{13}$C 
potential shown in Fig. \ref{pot1313}A. The main difference between 
the two figures is that the ground state of the compound nucleus 
$^{25}$Mg is located at a higher energy than $^{26}$Mg
because of the unpaired neutron in $^{25}$Mg.

\begin{figure}
\includegraphics[width = 10cm]{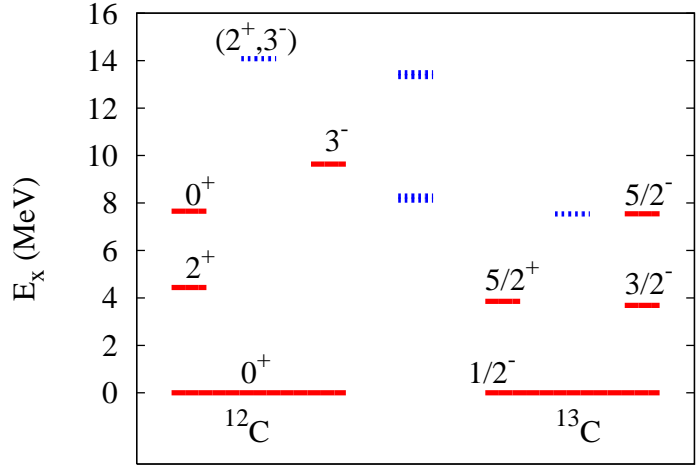}
\caption{\label{spc1213} 
(Color online) 
Excitation energies of the channels considered in the calculations 
of the $^{12}$C+$^{13}$C fusion cross section.
The (red) solid lines are the states given in Table I.
The blue lines are mutual excitations considered in the calculations.}
\end{figure}

The excitation energies that are considered in the calculations 
of the fusion cross sections for $^{12}$C+$^{13}$C are shown in 
Fig. \ref{spc1213}. The solid (red) lines represent the excitations
shown in Table I which together with the elastic channel give a 
total of 7 channels. One of these states is the $0_2^+$ Hoyle state
of $^{12}$C which is included because its coupling to the $2^+$ is
known.
 
The (blue) dashed lines in Fig. \ref{spc1213} are the 6 mutual 
excitations that are considered. They include the $(2^+,3^-)$
mutual excitation in $^{12}$C, the $(3/2^-,5/2^+)$ mutual
excitation in $^{13}$C, and the four $(I_1,I_2)$ mutual excitations
in projectile and target with $I_1$ = $2^+$ or $3^-$ and $I_2$ = 
$3/2^-$ or $5/2^+$. Two-phonon excitations (except the $0_2^+$
Hoyle state which has already been mentioned) and mutual 
excitations that involve the $5/2^-$ state in $^{13}$C are ignored. 
The total number of excitation channels is therefore 7+6 = 13. 
When the one-neutron transfer with zero Q value is included in the 
calculations as described in Sect. II.D, the number of channels is
doubled. The full calculation, which includes the one-neutron 
transfer and all of the 13 excitation channels shown in Fig. 
\ref{spc1213} will therefore have 26 channels and it is referred 
to as Ch26.

A major problem, which causes some uncertainty in the
predictions made by the coupled-channels calculations, is the high 
excitation energy and the large strength of the octupole transition 
in $^{12}$C (see Table 1.)
A large excitation energy implies decaying boundary conditions at 
large separations of projectile and target instead of the scattering 
boundary conditions, and this can cause numerical problems. 
A large octupole strength implies a strong sensitivity to mutual 
excitations that involve the $3^-$ state and some of these excitations 
are poorly known as explained below.
The octupole excitation of $^{13}$C, on the other hand, is not a problem
because the excitation energy of the $5/2^+$ state is much lower than in
$^{12}$C, and the effective octupole strength is reduced as shown in Eq. 
(\ref{octupole}).

\begin{figure}
\includegraphics[width = 10cm]{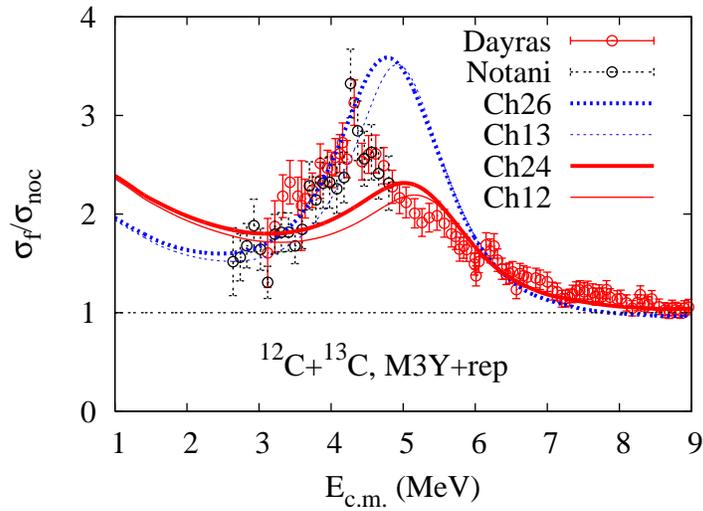}
\caption{\label{frat1213} 
(Color online) 
Enhancement factors of the measured $^{12}$C+$^{13}$C fusion cross sections 
\cite{dayras,tang1} relative to the cross sections obtained in the 
no-coupling limit.  The curves show the enhancement of calculations 
that include the 24 (Ch24) and 26 (Ch26) channels described in the text.
The associated calculations without transfer, Ch12 and Ch13, are shown
by the thin curves.}
\end{figure}

The enhancement of the $^{12}$C+$^{13}$C fusion relative to the 
no-coupling limit is shown in Fig. \ref{frat1213} for different 
coupled-channels calculations. 
The full calculation Ch26 has a very strong peak near 4.7 MeV. 
The calculation Ch24, which excludes the mutual $(2^+,3^-)$ 
excitation in $^{12}$C, has a much weaker peak.
There are two reasons why the mutual $(2^+,3^-)$ excitation 
in $^{12}$C has such a large influence on subbarrier fusion. 
One reason is the large octupole strength of $^{12}$C.
The other reason is that the direct coupling of the
ground state to mutual excitations is governed by the second 
derivative of the ion-ion potential, according to Eq. (\ref{uexp}),
and this quantity is particularly large for the M3Y+repulsion 
potential as pointed out in section III.C.

It is very interesting that the enhancement of the data \cite{dayras,tang1}
with respect to the no-coupling limit also exhibits a peak structure
in Fig. \ref{frat1213}. The peak is located near 4.3 MeV, somewhat 
below the positions of the two calculated peaks. 
The influence of transfer can be seen by comparing the thin curves 
C12 and Ch13, which do not include the influence of transfer, to the 
corresponding thick curves Ch24 and Ch26, which include the effect. 
The influence is relatively modest but it does shift the calculated 
peaks towards the peak position of the data.

\begin{figure}
\includegraphics[width = 10cm]{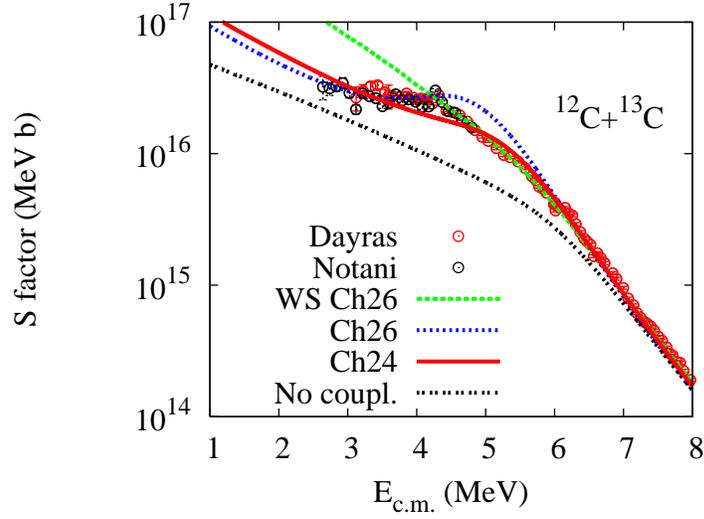}
\caption{\label{fsf1213} 
(Color online) 
$S$ factors for the fusion of $^{12}$C+$^{13}$C \cite{dayras,tang1}
are compared to calculations that are based on a Woods-Saxon 
potential and 26 channels (WS Ch26), and on the M3Y+repulsion 
potential with 24 and 26 channels, respectively.
The no-coupling limit is also shown.} 
\end{figure}

The $S$ factor for the fusion of $^{12}$C+$^{13}$C is illustrated 
in Fig. \ref{fsf1213}.  The calculation `WS Ch26' is based on the 
standard Woods-Saxon potential with radius $R_{ws}$ = 5.352 fm, 
depth $V_0$ = -38.6 MeV, and diffuseness $a$ = 0.63 fm, which was 
illustrated in Fig. \ref{pot1313}B. It includes all 26 channels
considered in this sections and makes a very good fit to 
the data above 4 MeV. However, a strong hindrance of the data 
sets in at energies below 4 MeV. The comparison shows that the 
fusion of $^{12}$C+$^{13}$C is an example on the fusion
hindrance phenomenon which has been observed in many medium 
heavy systems \cite{niy,sys}. 

The two calculations Ch24 and Ch26 that are based on the 
M3Y+repulsion potential reproduce the data in Fig. \ref{fsf1213} 
at the lowest and also at high energies. The calculation Ch26 
exceeds the data between 4.5 and 5.5 MeV by up to 50\%. 
The calculation Ch24 is in better agreement with the data but 
there are still some deviations.
The measured $S$ factor is rather flat between 3 and 4 MeV and 
has a modest peak near 4.3 MeV. 
The calculated $S$ factors do not exhibit any sharp peak 
structures but rise slowly with decreasing energy.

The two calculations Ch24 and Ch26 represent extreme views of 
the influence of the mutual excitation of the $2^+$ and $3^-$ 
states in $^{12}$C, i.~e., of a quadrupole excitation built on
the $3^-$ state or an octupole excitation built on the $2^+$
state. The calculation Ch24 completely ignores it, whereas the 
calculation Ch26 exaggerates it by assuming that the $2^+$ and 
$3^-$ excitations are independent modes of excitation. 
In a realistic, microscopic description,
the mutual excitation of the $2^+$ and $3^-$ states will not be 
one discrete excitation but will be fragmented and spread out 
over a range of excitation energies.
This is indeed the result of shell model calculations
that are discussed in section VI.

\section{Predictions of the fusion of $^{12}$C+$^{12}$C}

Similar predictions are made for the fusion of $^{12}$C+$^{12}$C.
The entrance channel potential that is used is shown in Fig. 
\ref{pot1313}C. It was obtained with the parameters of the 
point-proton density of $^{12}$C shown in Table I, and the 
diffuseness associated with the repulsive part of the M3Y+repulsion 
interaction was again set equal to the value $a_r$ = 0.31 fm which 
was determined in the analysis of the $^{13}$C+$^{13}$C fusion data.

\begin{figure}
\includegraphics[width = 10cm]{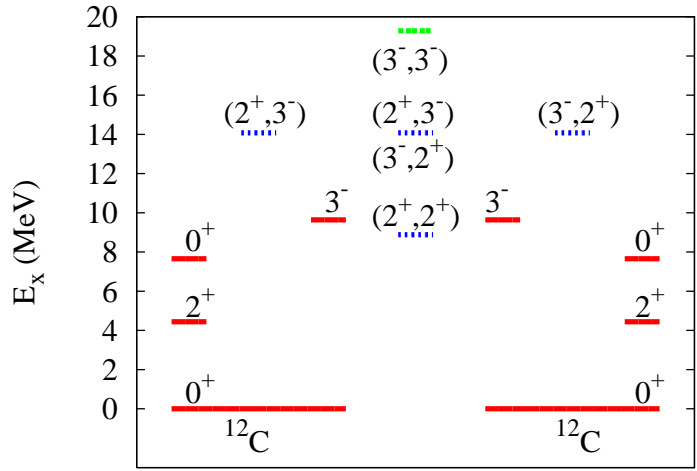}
\caption{\label{spc1212} 
(Color online) 
Excitation energies of the channels considered in the calculations
of the $^{12}$C+$^{12}$C fusion cross section. The (red) solid 
lines are the states given in Table I.
The (blue) dashed lines are the mutual excitations considered in the
calculations.  The $(3^-,3^-)$ mutual excitation of projectile and
target is indicated but it is ignored in the calculations.}
\end{figure}

The excitation energies that are considered in the coupled-channels
calculations are shown in Fig. \ref{spc1212}. The solid (red) lines 
represent the excitations shown in Table I which together with 
the elastic channel give a total of 7 channels. 
The second $0^+$ state is Hoyle state.
The (blue) dashed lines in Fig. \ref{spc1212} are the mutual 
excitations that are considered in the calculations. They include 
the three mutual excitations: $(2^+,2^+)$, $(2^+,3^-)$ and 
$(3^-,2^+)$, with a one-phonon excitation in each $^{12}$C nucleus, 
and the two mutual excitations: $(2^+,3^-)$, with both one-phonon
excitations belonging to the same nucleus. 
There is also a $(3^-,3^-)$ mutual excitation indicated in the
figure (with one octupole excitation in each nucleus) but it is 
ignored because the excitation energy is very high. 
The total number of channels consisting of the 7 channels 
mentioned above (represented by the solid (red) lines
in Fig. \ref{spc1212}) and the 5 mutual excitation channels 
(the (blue) dashed lines in Fig. \ref{spc1212}) is therefore 12.
The associated coupled-channels calculation is referred to as Ch12.

\begin{figure}
\includegraphics[width = 12cm]{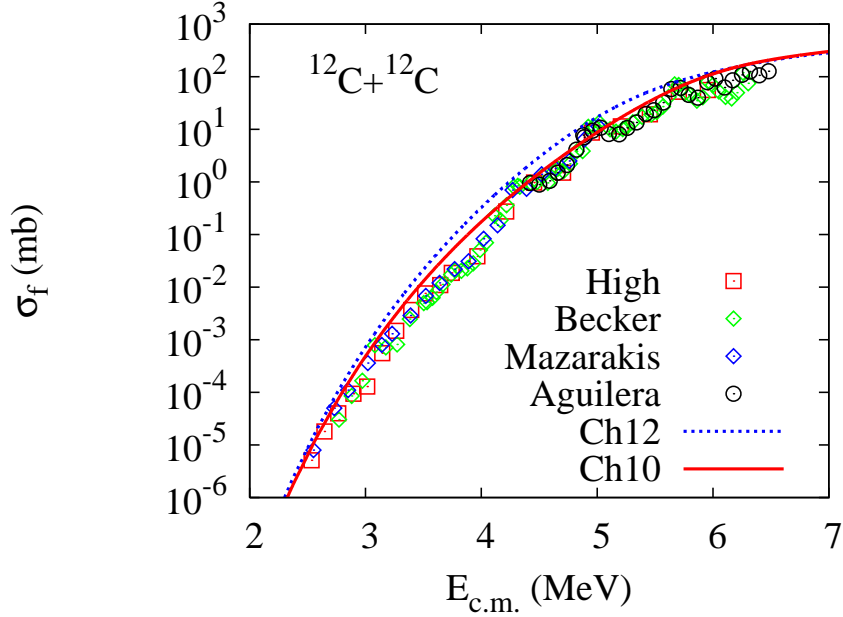}
\caption{\label{ff1212}
(Color online) 
Measured fusion cross sections for $^{12}$C+$^{12}$C 
\cite{aguilera,becker,high,maza} 
are compared to the Ch10 and Ch12 coupled-channels calculations
discussed in the text.
The data of Refs. \cite{maza,high} have been shifted in energy
($\Delta E$ = +100 keV \cite{maza} and 
$\Delta E$ = +75 keV \cite{high}, respectively) following the
suggestions of Ref. \cite{aguilera}.}
\end{figure}

The calculation Ch12 includes the mutual excitation of the $2^+$ 
and $3^-$ states, not only when the two states belong to different 
nuclei but also when they belong to the same nucleus. 
Since the description of the $(2^+,3^-)$ mutual excitation in
terms of two independent modes of excitation is questionable
when the two excitations belong to the same nucleus, it is of interest 
to perform a calculation that does not include that kind of mutual
excitation. Such a calculation contains 10 channels and is
referred to as Ch10. The Ch10 calculation does contain mutual 
excitations of the $2^+$ and $3^-$ states but only when the 
two one-phonon excitations belong to different nuclei.
The two calculations are compared in Fig. \ref{ff1212}
to the measured cross sections of Ref. 
\cite{maza,high,aguilera,becker}. 
The Ch12 calculation exceeds the data substantially between 3 and 
5.5 MeV, whereas the Ch10 calculation is lower and intersects with
some of the measured peak cross sections.

\begin{figure}
\includegraphics[width = 12cm]{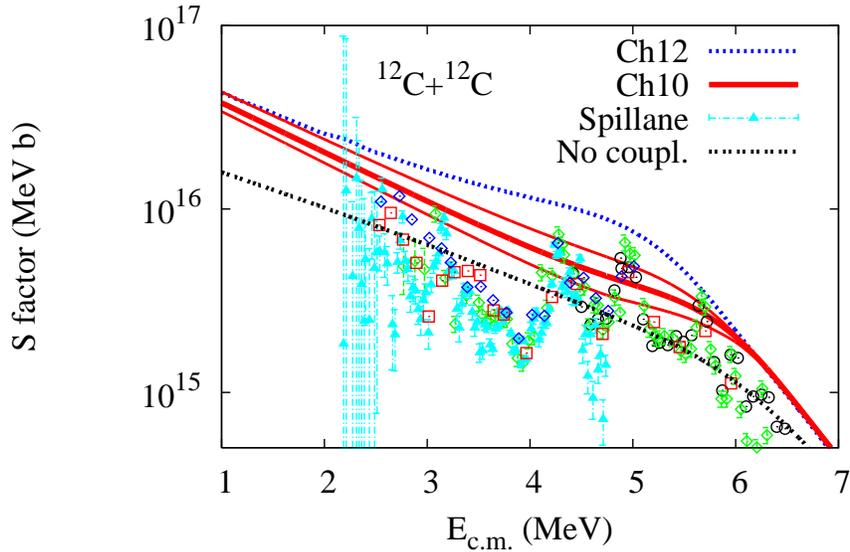}
\caption{\label{fsf1212}
(Color online) 
Measured $S$ factors for the fusion of $^{12}$C+$^{12}$C 
\cite{aguilera,high,maza,becker,spillane} 
are compared to the coupled-channels calculations Ch10 and Ch12, 
and to the no-coupling limit. The thin solid curves show the error 
band in Ch10 calculations due to the uncertainty in the octupole 
strength, $\beta_3$ = 0.90 $\pm$ 0.07.}
\end{figure}

The $S$ factors for the fusion of $^{12}$C+$^{12}$C predicted by the 
Ch10 and Ch12 calculations are compared in Fig. \ref{fsf1212} to the 
data of five experiments \cite{maza,high,aguilera,becker,spillane}.
The symbols for the four measurements \cite{maza,high,aguilera,becker}
are the same as in Fig. \ref{ff1212}, whereas the symbol for
Spillane's measurements \cite{spillane} is indicated in the figure.
The calculation Ch12 exceeds all of the data points shown in this 
figure. It is considered an extreme upper limit because it exaggerates,
as discussed above and in the previous section, the influence of the 
mutual $(2^+,3^-)$ excitation within the same nucleus.  The calculation 
Ch10, which is much lower, is considered a lower limit of the prediction.


The large difference between the Ch10 and Ch12 calculations shown
in Fig. \ref{fsf1212} is very unfortunate. It illustrates the
difficulty and uncertainty in predicting the fusion cross section 
due to the poor empirical knowledge of excitations that are built 
on the $2^+$ and $3^-$ states. We shall eliminate that problem 
in the next section by applying the results of shell model
calculations. Another uncertainty, which is similar in magnitude,
is due to the empirical value $\beta_3$ = 0.90(7) \cite{ENDSF}. 
This is illustrated in the figure by the thin lines and represents 
an error band of up to $\pm$25\%.


The $S$ factor one obtains in the no-coupling limit is shown by the
lowest dashed curve in Fig. \ref{fsf1212}. 
It is a factor of 2 to 3 below the measured peak cross sections 
and a factor of 2 to 5 above the minima of the measured cross 
sections. In other words, it does not provide a background nor 
does it provide an upper limit of the measured cross sections.
It is closer to the smooth experimental cross section constructed 
by Yakovlev et al. \cite{yakovlev}.
%

\section{Systematics of the fusion of carbon isotopes}

The discussion in the previous sections shows that it is possible to 
develop a fairly comprehensive description of the fusion data for the 
three carbon systems within the coupled-channels approach.
A disturbing trend in the analysis of the $^{13}$C+$^{13}$C 
fusion data \cite{trent} is the need for dividing the data by the factor 
$S_c$ = 0.843, i.~e., for increasing the measured cross sections by 19\%.
However, that is not a very serious problem because the systematic error
of the experiment \cite{trent} is 15\%.

\begin{figure}
\includegraphics[width = 12cm]{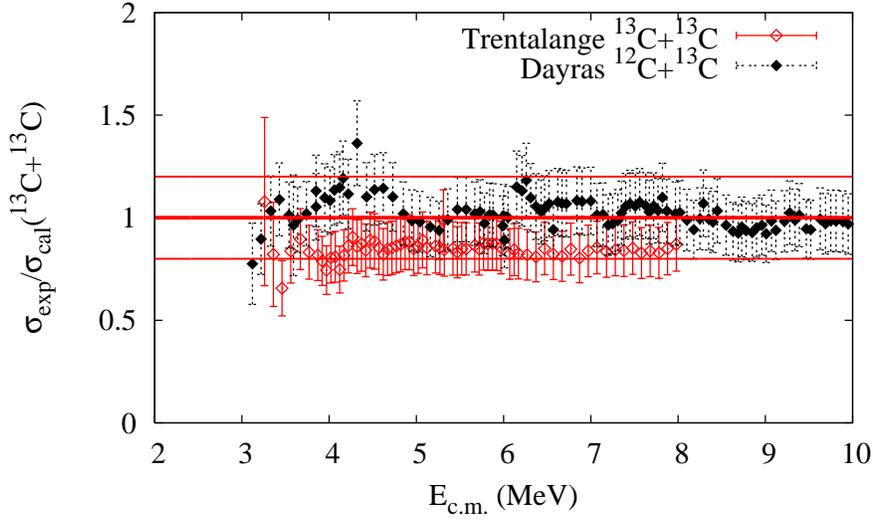}
\caption{\label{frfcall} (Color online) 
Ratios of measured fusion cross sections relative to the calculation 
Ch26 of the fusion of $^{13}$C+$^{13}$C. The error bars include a 15\%
systematic error. 
The average ratio for the $^{12}$C+$^{13}$C data \cite{dayras}
is 1.02, whereas the average ratio for the $^{13}$C+$^{13}$C data
\cite{trent} is 0.843.}
\end{figure}

The need for a renormalization of the $^{13}$C+$^{13}$C data is possibly 
an experimental problem because it is difficult to explain why the 
measured fusion cross sections for $^{12}$C+$^{13}$C \cite{dayras} are
larger than the measured cross sections for $^{13}$C+$^{13}$C \cite{trent}.
Naively, one would expect the cross sections for the smaller system to be 
smaller. This is indeed the ordering that is observed in the fusion of 
carbon isotopes with a thorium target \cite{alcorta}.
Of course, the problem could be in the absolute normalization of 
$^{12}$C+$^{13}$C data \cite{dayras}. Whatever the explanation is, the 
basic problem is illustrated in Fig.  \ref{frfcall}, where the measured 
cross sections for the two systems have both been normalized to the Ch26 
calculation for the $^{13}$C+$^{13}$C system discussed in sections III.B. 
It is seen that the normalized cross sections are fairly constant but 
the cross sections for the smaller $^{12}$C+$^{13}$C system are
larger, on average by the factor 1.02/0.84 = 1.21 
(c.~f. the caption to Fig. \ref{frfcall}). 
The 21\% deviation between the two measurements is within the large 
uncertainties of the two experiments, each being of the order of 15 to 
30\% but it does show an unexpected trend.  Clearly, it is desirable
to have the experimental uncertainties reduced in future measurements.

Another disturbing feature discussed in the previous section is the 
large uncertainty in the cross section predicted for the fusion of 
$^{12}$C+$^{12}$C. Half of the uncertainty concerns the influence 
of the $(2^+,3^-)$ mutual excitation which is poorly known 
experimentally \cite{ENDSF}. One way to overcome this problem is to 
rely on shell model calculations \cite{brown} which predict that E2 
excitations built on the $3^-$ state and E3 excitations built on 
the $2^+$ state primarily populate a $4^-$ state which is about 
4 MeV above the $3^-$ state. That is similar to the model we have 
used for the mutual $(2^+,3^-)$ excitation. However, the B-values of 
these transitions are only about half of the B-values obtained 
for the corresponding E2 and E3 transitions from the ground state 
to the first $2^+$ and $3^-$ states in $^{12}$C.

\begin{figure}
\includegraphics[width = 12cm]{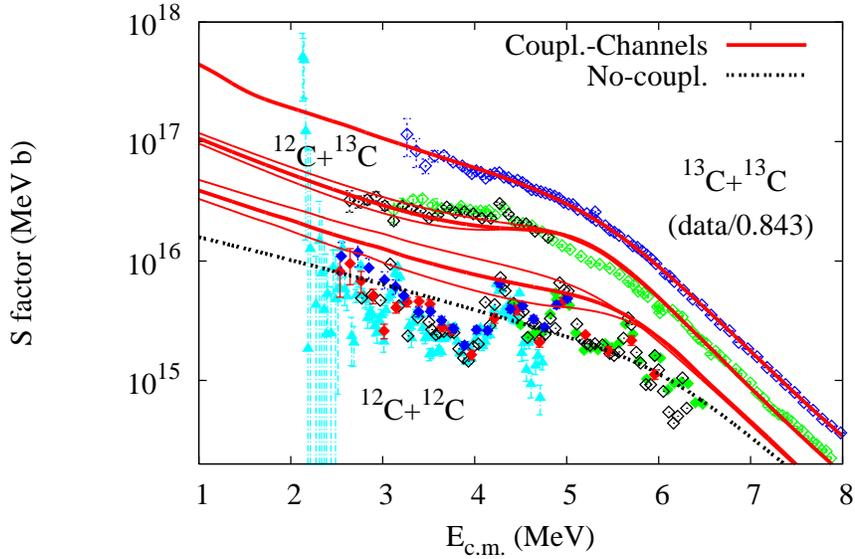}
\caption{\label{fsfcallsm} (Color online) 
$S$ factors for the fusion of 
$^{13}$C+$^{13}$C \cite{trent}, $^{12}$C+$^{13}$C \cite{dayras,tang1} and 
$^{12}$C+$^{12}$C \cite{aguilera,high,maza,becker,spillane} 
are compared to the coupled-channels calculations described in the text.
The $^{13}$C+$^{13}$C data have been divided by $S_c$ = 0.843 which 
optimizes the fit to the data. 
The thin solid curves show the error band due to the uncertainty in the 
octupole strength in $^{12}$C, $\beta_3$ = 0.90 $\pm$ 0.07.
The black dashed curve is the no-coupling limit for $^{12}$C+$^{12}$C.} 
\end{figure}

The cross sections one obtains for the three fusion reactions 
discussed in this work, using the best fit parameters determined 
in the previous sections, and the shell model results described 
above, are compared to the data in Fig. \ref{fsfcallsm} in terms 
of the $S$ factor.
The data for $^{13}$C+$^{13}$C have been divided by the scaling 
factor $S_C$ = 0.843 which results in an excellent 
agreement with the data.
The calculations of the fusion of $^{12}$C+$^{12}$C and $^{12}$C+$^{13}$C 
are Ch12 and Ch26 calculations, respectively, which are similar to those
discussed in section IV and V. The only difference is that the parameters 
for the mutual $(2^+,3^-)$ excitation in $^{12}$C have been replaced by 
shell model predictions \cite{brown}.

The calculated $S$ factors for the fusion of $^{12}$C+$^{12}$C and 
$^{12}$C+$^{13}$C are essentially predictions because the parameters 
of ion-ion potential and the density of $^{13}$C were determined in 
the analysis of the $^{13}$C+$^{13}$C fusion data, whereas the density 
of $^{12}$C was determined by a fit to the experimental point-proton 
distribution.  The $^{12}$C+$^{13}$C data \cite{dayras,tang1} shown 
in Fig.  \ref{fsfcallsm} form a plateau between 2.5 and 4 MeV and has 
a peak near 4.3 MeV.
These features are not quite reproduced by the calculation but the 
fit to the data is not poor; the fit to the data of Ref. \cite{dayras}
is stable with respect to the scaling factor $S_c$ and to an overall
energy  shift $\Delta E$ in the center of mass energy, i.~e.,
the $\chi^2$ per data point has a minimum for $S_c$ = 1 and
$\Delta E$ = 0.

Although there is still some  uncertainty in the predicted $S$ factor
for the fusion of $^{12}$C+$^{12}$C, it appears that the calculation 
shown in Fig. \ref{fsfcallsm} is consistent with the maxima 
of the measured peak cross sections. 
The peak at 4.924 MeV is within the error band of the calculation.
The 3 low-energy (triangular shaped) data points are from 
Ref. \cite{spillane}. They exceed the calculation tremendously but 
that is not a problem either because those 3 data points are 
questionable as discussed in Ref. \cite{sisyfus,tang2}.
The calculation (with its error band) is therefore expected to provide 
a useful guidance and an upper limit for the extrapolation to the low 
energies that are of interest to astrophysics.

\section{Conclusions}

We have explored the fusion of different combinations of carbon isotopes 
within the coupled-channels approach. 
The fusion cross section is determined by ingoing-wave-boundary
conditions that are imposed at the minimum of the pocket in the
entrance channel potential. 
It turned out that the measured cross sections for the fusion of 
$^{13}$C+$^{13}$C cannot be explained accurately at low energies
when a standard Woods-Saxon potential is applied in the 
calculations.  However, the data can be explained fairly well by applying 
the shallow potential that is produced by the M3Y+repulsion potential. 
This is a characteristic feature of the fusion hindrance phenomenon 
which has been observed in many medium-heavy systems.
One reason the shallow potential makes it possible to explain the 
carbon data by coupled-channels calculations is that the high-lying 
excitations become closed channels at low center-of-mass energies, 
not only with respect to reactions but also to fusion.

The coupled-channels calculations are surprisingly sensitive to mutual 
excitations and this seems to be justified by the comparison to the data.
Thus the calculations that include one-phonon excitations alone cannot 
explain the data so well, neither when the standard Woods-Saxon nor 
the M3Y+repulsion potential is applied. 
The data are reproduced much better when mutual excitations are 
included and the shallow M3Y+repulsion potential is applied.
The reason is that the direct coupling to these high-lying states is
proportional to the second derivative of the ion-ion potential and this
quantity is much larger for the M3Y+repulsion than for the standard
Woods-Saxon potential because of the strong repulsion. 

The large sensitivity to the excitation of high-lying states is 
particularly critical when the mutual excitations involve the 
strong octupole excitation of $^{12}$C. 
The uncertainty in the strength of the octupole excitation, and 
also in the strength of transitions to high-lying states that are 
built on the $2^+$ and $3^-$ states, makes it difficult to predict 
the fusion cross sections very accurately when $^{12}$C is involved. 
However, by adopting the nuclear structure properties of the 
high-lying states predicted by the shell model, one obtains a 
fairly good description of the $^{12}$C+$^{13}$C fusion data. 

The prediction of the cross section for the fusion of $^{12}$C+$^{12}$C,
based on the ingoing-wave-boundary conditions and the shell model
prediction of the coupling to high-lying states, 
exceeds most of the measured cross sections and is consistent with 
the maxima of the observed peak cross sections.
Thus it appears that the calculation provides an upper limit of the 
measured cross sections. 
The calculation is therefore expected to provide an upper limit for 
the extrapolation of cross sections into the unexplored territory of 
very low energies.

On the experimental side it is very important to get a better 
absolute normalization of the measured fusion cross sections. 
The current systematic errors are fairly large. This was exploited 
in the analysis of the $^{13}$C+$^{13}$C fusion data by focusing 
on the shape of the measured cross section, whereas the absolute 
normalization was treated as an adjustable parameter. 
In this way it was possible to calibrate the M3Y+repulsion potential 
and achieve an excellent fit to the data. It is of great interest to 
know whether the renormalization that was applied to the data can 
be justified by more accurate measurements. 

{\bf Acknowledgments}. 
The authors are grateful to B. A. Brown for providing the shell model
calculations. This work was supported by the U.S. Department of Energy, 
Office of Nuclear Physics, contract no. DE-AC02-06CH11357.
One of us (X.T.) is supported by the NSF under Grant No. PHY-0758100
and PHY-0822648, and the National Natural Science Foundation of
China under Grant No. 11021504, and the University of Notre Dame.


\section{Appendix: Couplings}

The matrix elements of the couplings between two heavy ions are
generated by matrix elements of the static or dynamic deformation 
amplitudes $\alpha_{\lambda\mu}$ (see for example Ref. \cite{alge}).
The matrix elements are expressed in terms of the reduced
matrix elements $\langle||\alpha_\lambda||\rangle$,
\begin{equation}
\langle I_2M_2  |\alpha_{\lambda\mu} | I_1M_1 \rangle =
\langle I_1M_1 \lambda\mu|I_2M_2\rangle \
\frac{\langle I_2||\alpha_\lambda ||I_1 \rangle}
{\sqrt{2I_2+1}}.
\label{matrixelm}
\end{equation}
The off-diagonal reduced matrix elements can be extracted from 
the measured $B(E\lambda)$ values according to the formula 
\begin{equation}
B(E\lambda, I_2\rightarrow I_1) =
\bigl(\frac{3Ze^2R_C^\lambda}{4\pi}\bigr)^2 \
\frac{|\langle I_2||\alpha_\lambda || I_1\rangle|^2}{2I_2+1}.
\end{equation}
If the $B$-value is expressed in Weisskopf units the relation is
\begin{equation}
B(E\lambda)_{W.u.} =
\frac{[Z(\lambda+3)]^2}{4\pi}
\frac{|\langle I_2||\alpha_\lambda || I_1\rangle|^2}{2I_2+1}.
\end{equation}
The diagonal matrix elements, on the other hand, can be obtained
from the measured quadrupole moments,
\begin{equation}
Q_{II} = \sqrt{\frac{16\pi}{5}} \ 
\frac{3ZeR_C^2}{4\pi} \
\langle II |\alpha_{20}| II\rangle.
\end{equation}
They are related to the intrinsic quadrupole moments $Q_0$ 
and the $K$ quantum number by
\begin{equation} Q_{II} = \langle IK 20| IK\rangle \
\langle II 20| II\rangle \ Q_0.
\end{equation}

For a deformed nucleus with a static deformation $\beta_\lambda$
the deformation amplitude is $\alpha_{\lambda\mu}$ = 
$\beta_\lambda D^{\lambda}_{\mu 0}({\hat e})$,
where ${\hat e}$ is the orientation of the symmetry axis.
In this case the expressions for the matrix elements 
between states $|IKM\rangle$ are
\begin{equation}
\langle I_2KM|\alpha_{\lambda0}| I_1KM\rangle =
\beta_\lambda \sqrt{\frac{2I_1+1}{2I_2+1}} \
\langle I_1M \ \lambda 0| I_2M\rangle \
\langle I_1K \ \lambda 0| I_2K\rangle.
\label{matel}
\end{equation}
This implies (by comparing to Eq. (\ref{matrixelm}) that the 
reduced matrix elements are
\begin{equation}
\langle I_2||\alpha_\lambda ||I_1 \rangle
= \beta_2 \sqrt{2I_1+1} \ 
\langle I_1K \ \lambda 0| I_2K\rangle.
\end{equation}
\begin{equation}
B(E\lambda)_{W.u.} =
\frac{[Z(\lambda+3)]^2}{4\pi} \ \beta_\lambda^2 \
\frac{2I_1+1}{2I_2+1} \
\langle I_1K \ \lambda 0| I_2K\rangle^2.
\label{belwu}
\end{equation}

\begin{table}
\caption{Parameters of the densities 
that are used in calculating the M3Y+repulsion potential.
The second column shows the radius $R$; the diffuseness 
was kept fixed at $a$ = 0.44 fm and the third column is the calculated
rms radius. Also shown is the diffuseness $a_r$ that is used in calculating 
the repulsive part of the potential, the value of scaling factor $S_c$ 
that optimizes the fit to the $^{13}$C+$^{13}$C fusion data  
\cite{trent}, and the $\chi^2/N$. 
The last 3 lines show the radii that are consistent with the rms radii of
the point-proton (pp) and matter distributions of $^{12}$C and $^{13}$C.} 
\begin{ruledtabular}
\begin{tabular} {|c|c|c|c|c|c|} 
Nucleus & R (fm) & rms (fm) & $a_r$ (fm) & $S_c$ & $\chi^2/N$ \\ 
\colrule
$^{13}$C  & 2.17  & 2.345 & 0.33 & 1     & 2.75 \\ 
$^{13}$C  & 2.28  & 2.407 & 0.31 & 0.843 & 1.00  \\ 
\colrule
$^{12}$C (pp)   & 2.155 & 2.337(2) & & & \\ 
$^{13}$C (pp)   & 2.146 & 2.332(3) & & & \\ 
$^{13}$C, Eq. (\ref{rms}) & 2.228 & 2.378    & & & \\
\end{tabular}
\end{ruledtabular}
\end{table}
\begin{table}
\caption{Properties of E2 and E3 transitions 
in $^{12}$C and $^{13}$C \cite{ENDSF}.
The intrinsic quadrupole moments have been extracted from the lowest
quadrupole transitions.}
\begin{ruledtabular}
\begin{tabular} {|c|c|c|c|c|c|}
Nucleus & State & E$_x$ (MeV) & Transition & 
B(E$\lambda$) (W.u.) &
$\beta_\lambda^C$ \\
\colrule
$^{12}$C & $0^+_1$  &  0  &     & $Q_0$ =-19.5 fm$^2$ & 0.570 \\ 
 & $2^+$   & 4.439 & E2: $0_1^+\rightarrow2^+$ & 4.65(26)  & 0.570 \\ 
 & $0_2^+$ & 7.654 & E2: $2^+\rightarrow0_2^+$ & 8.0(11)   & 0.236 \\ 
 & $3^-$   & 9.641 & E3: $0_1^+\rightarrow3^-$ & 12(2) \cite{ENDSF} & 0.90(7) \\ 
\colrule
$^{13}$C & $1/2^-$   &  0 &  & $Q_0$ = -17.9 fm$^2$ & 0.495    \\
 & $3/2^-$ & 3.6845 & E2: $1/2^-\rightarrow3/2^-$ & 3.5(8) & 0.495 \\
 & $5/2^-$ & 7.547  & E2: $1/2^-\rightarrow5/2^-$ & 3.1(2) & 0.465 \\
 & $5/2^+$ & 3.8538 & E3: $1/2^-\rightarrow5/2^+$ & 10(4)  & 0.82  \\ 
\end{tabular}
\end{ruledtabular}
\end{table}


%


\end{document}